\documentclass[conference]{IEEEtran}
\usepackage{comment}
\usepackage{booktabs}
\usepackage{stfloats}
\usepackage{bm}
\usepackage{graphicx}
\usepackage{amssymb}
\usepackage{amsfonts}
\usepackage{amsmath}
\usepackage{subfigure}
\usepackage{epstopdf}
\newtheorem{thm}{Theorem}

\newcommand\dif{\mathrm{d}}

\begin{document}
%
\title{On the Power Allocation for Hybrid DF and CF Protocol with Auxiliary Parameter in Fading Relay Channels}
\author{\authorblockN{Zhengchuan Chen\authorrefmark{1}\authorrefmark{2}, Pingyi Fan\authorrefmark{1}, Dapeng Wu\authorrefmark{2}  and Liquan Shen\authorrefmark{2}\authorrefmark{3}\\
\authorblockA{\authorrefmark{1}Tsinghua National Laboratory for Information
Science and Technology, and \\
Department of Electronic Engineering,
Tsinghua University, Beijing, China.\\
\authorrefmark{2}Department of Electrical and Computer Engineering,
University of Florida, Gainesville, FL 32611.\\
\authorrefmark{3}School of Information Science and Technology,
Shanghai University, Shanghai, China.}
E-mails:\authorrefmark{1}\{chenzc10@mails, fpy@\}.tsinghua.edu.cn,
\authorrefmark{2}wu@ece.ufl.edu,\authorrefmark{3}jsslq@shu.edu.cn } } \maketitle
\begin{abstract}
In fading channels, power allocation over channel state may bring a
rate increment compared to the fixed constant power mode. Such a
rate increment  is referred to power allocation gain. It is expected
that the power allocation gain varies for different relay protocols.
In this paper, Decode-and-Forward (DF) and Compress-and-Forward (CF)
protocols are considered. We first establish a general framework for
relay power allocation of DF and CF over channel state in
half-duplex relay channels and present the optimal solution for
relay power allocation with auxiliary parameters, respectively.
Then, we reconsider the power allocation problem for one hybrid
scheme which always selects the better one between DF and CF and
obtain a near optimal solution for the hybrid scheme by introducing
an auxiliary rate function as well as avoiding the non-concave rate
optimization problem. 
\end{abstract}

%

\section{Introduction}

In cooperative communication networks, power allocation over channel
state may bring rate gains \cite{Thomas}. However, it is not easy to find the
optimal power allocation because the exact capacity of most of the
wireless networks has not been known. There were some useful
cooperation strategies put forward in the literatures which provided
efficient approach to transmit information and gave lower bounds for
the rate performance of the system. For instance, two relay
protocols, Decode-and-Forward (DF) and Compress-and-Forward (CF),
were proposed in \cite{Gamal} to evaluate the information rate for
relay channels (RC). Particularly, the DF protocol was shown to be
able to achieve the capacity of degraded RC \cite{Gamal} and sender
frequency division RC \cite{Sina}. Due to the effectiveness of DF
and CF, they has been widely used in cooperative communication
networks, achieving good rate performance in various networks
\cite{Kramer}\cite{Kim}.

Based on DF and CF protocols, power allocation can be naturally
extended to general networks to combat the varying channel states.
Take the RC as an example again. Since there are only three wireless
links in the system, it is available for the source and the relay to
know the current channel gains before transmissions via timely
feedback from the receiver. The global power allocation over fading
channel problem in RC has been studied in \cite{Ander}. By assuming
that the source and the relay subject to a sum power constraint, the
authors provided algorithms on how to find the optimal power
allocation. It is also noted that the power allocation established
in \cite{Ander} achieved the maximal throughput of the relay-receive
phase and relay-transmit phase in half-duplex relay channels (HDRC).
The result was implicitly based on a buffer at the relay such that
if the relay-destination channel is worse, it can store the message
and transmit them when the relay-destination channel becomes better.
In practice, if the relay has a finite storage and limited
processing capability, the system may become unstable and the power
allocation gain will degrade.

To improve the achievable rate, selecting better relay protocol
among multiple protocols provides another alternative. This
intuition comes from theoretical analysis on combining DF and CF in
static RC \cite{Che}-\cite{Chen2}. It was found superposition structure of the DF and CF
codewords provides some rate gain with penalties of decoding
complexity \cite{Gamal}\cite{CM}. Moreover, a general insight was
also obtained that DF outperforms CF for only some of the channel
gain combinations while the relationship reverses for the others.
This implies that in fading relay channels, a hybrid scheme which
selects the better one between DF and CF according to the channel
state may provide some rate gains while avoiding the complicated
codeword design.

Instead of using other techniques, e.g., \cite{Zhang}-\cite{Chan}, to combat channel fading, in this work, we thoroughly analyze the relay
power allocation over channel state when the relay adopts both DF
and CF protocols. 

The remainder of this paper is organized as follows. In Section II,
we introduce the system model and establish a general framework for
the relay power allocation problem. In Section III, we present a
parameterized form solution for the problem corresponding to DF and
CF, respectively. In Section IV, we further investigate the relay
power allocation corresponding to the hybrid scheme and discuss the
optimal solution by introducing an auxiliary rate function.

\section{System model and problem preliminary}
\begin{figure}[!t]
\centering
\includegraphics[width=0.5\textwidth]{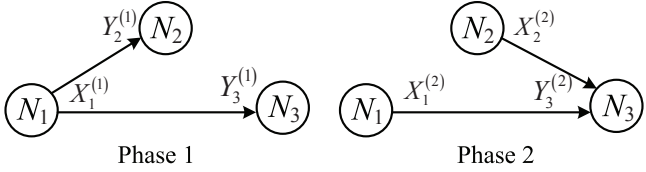}
\caption{The two-phase transmission of a half-duplex relay channel.}
\label{relay-ch}
\end{figure}

Let us consider a HDRC as illustrated in Fig.
\ref{relay-ch}. In the figure, $N_1$, $N_2$ and $N_3$ represent the
source, the relay and the destination, respectively. We assume
the relay is operated in half-duplex manner. Due to the multipath
effect, the channel gains are varying along with the time. We assume that the channel gains are
holding constant for a fixed time length which is referred to as a
block and the channel gains varies independently between consecutive
blocks. The signal transmissions in each block are divided into two
phases as depicted in Fig. \ref{relay-ch}. In \emph{Phase 1}, the
source transmits signal while the other two nodes listen. In
\emph{Phase 2}, the source and the relay transmit signals to the
destination. To distinguish the signal in different phases, let us
denote the complex baseband signal transmitted at $N_i~(i=1,2)$ and
received at $N_j~(j=2,3)$ in \emph{Phase $k$} $(k=1,2)$ by
$X_i^{(k)}$ and $Y_j^{(k)}$, respectively. For simplicity, we use
$H_{ji}$ and $h_{ji}$ to represent the channel gain variable
and its realization for $N_i$-$N_j$ link in each block. Accordingly,
transmissions in the HDRC can be expressed as
\begin{align}\label{e:channel1}
Y_3^{(1)}=&~h_{31}X_1^{(1)}+Z_3^{(1)}\\
Y_2^{(1)}=&~h_{21}X_1^{(1)}+Z_2^{(1)}\\
\label{e:channel3}Y_3^{(2)}=&~h_{31}X_1^{(2)}+h_{32}X_2^{(2)}+Z_3^{(2)}
\end{align}
where $Z_j^{(k)}~(j=2,3;~k=1,2)$ is additive white Gaussian noise
(AWGN) corresponding to $N_j$ in \emph{Phase $k$}. For simplicity,
we consider that the system is operated in unit bandwidth and we
assume that $Z_j^{(k)}$ obeys complex Gaussian distribution with
unit power spectrum density, i.e., $Z_j^{(k)}\sim\mathcal{CN}(0,1)$.
The source and the relay are assumed to know the channel gains at
the beginning of each block. In particular, as the channel
phase-shift is well-recovered at the receiver side, we focus on
$p_{_{|H_{ji}|}}(|h_{ji}|)$, the distribution of the amplitude of
$H_{ji}$.

For the reason of synchronization and power management, we assume
that the source transmit signal with the same power and the same
time length for the two phases. Denote the channel state by
$\vec{h}=(h_{31},h_{21},h_{32})$. By assuming that the block length
is long enough to support one time entire signalling, we can regard
the system as a static relay channel for each block. In general, in
the static case, the rate performance is a function of the receiver
side signal to noise ratio (SNR) of the three links. To focus on the
relay power allocation, we denote the receiver side SNR of
relay-destination link and the rate function in static HDRC by
$S_2\triangleq2|h_{32}|^2P_2$ and $R(S_2)$, respectively.

In fading HDRC, we consider long time average power constraint
$\overline{P_i}$ at $N_i~(i=1,2)$. Then the source transmits signal
with power $\overline{P_1}$ regardless of the channel state.
However, the relay can adjust $P_2$ adaptively w.r.t the channel
state $\vec{h}$ in each block. For clarity, we denote the relay
power allocation by $P_2(\vec{h})$.  The interest of this paper is
to find the optimal power allocation $P_2^{\star}(\vec{h})$
achieving the best rate performance of the system. Define
$S_2(\vec{h})\triangleq2|h_{32}|^2P_2(\vec{h})$. Regarding the
average rate as the measurement of the rate performance and taking
the average power constraint into consideration, we can specify the
relay power allocation problem as
\begin{align}
\label{e:ergodicdef}
\mathcal{P}:~~~~&\max_{S_2(\vec{h})}~~\left(\int_0^\infty\right)^3
p(\vec{h})R(S_2(\vec{h}))d\vec{h}\\
\label{e:cond2} &\text{~~s.t.} ~~~\left(\int_0^\infty\right)^3
p(\vec{h})\frac{S_2(\vec{h})}{2|h_{32}|^2}d\vec{h}=\overline{P_2},
\end{align}
where
$p(\vec{h})=p_{_{|H_{31}|}}(|h_{31}|)p_{_{|H_{21}|}}(|h_{21}|)p_{_{|H_{32}|}}(|h_{32}|)$;
$\left(\int_0^{\infty}\right)^3\triangleq
\int_0^{+\infty}\int_0^{+\infty}\int_0^{+\infty}$; and $\dif
\vec{h}\triangleq \dif |h_{31}|\dif |h_{21}|\dif |h_{32}|$.

If $R(S_2)$ is concave w.r.t. $S_2$, one can solve $\mathcal{P}$ by
Lagrangian method. Consider the Lagrangian
\begin{multline}
\nonumber
\mathcal{L}(S_2(\vec{h}),\mu)=\left(\int_0^\infty\right)^3p(\vec{h})R(S_2(\vec{h}))d\vec{h}\\
-\mu\bigg(\left(\int_0^\infty\right)^3p(\vec{h})\frac{S_2(\vec{h})}{2|h_{32}|^2}d\vec{h}-\overline{P_2}\bigg).
\end{multline}
Set $\frac{\partial\mathcal{L}(S_2(\vec{h}),\mu)}{\partial
S_2(\vec{h})}=0$. One has
\begin{align}\label{e:optcond2}
\frac{\dif R(S_2(\vec{h}))}{\dif
S_2(\vec{h})}-\frac{\mu}{2|h_{32}|^2}=0.
\end{align}
It should be
noted that only if the rate function $R(S_2)$ is concave w.r.t.
$S_2$ should the solution of \eqref{e:optcond2} be the optimal
$S_2(\vec{h})$.

\section{Optimal power allocation for DF and CF strategies}
In this section, we first analyze the concavity of DF rate and CF
rate. Then following necessary condition \eqref{e:optcond2}, we
present the optimal power allocation for DF and CF based on the
inverse function of the derivation of the rates.

\subsection{Concavity
of the DF rate and CF rate}
The rate achieved by DF protocol and CF protocol with Gaussian
signaling were presented in Proposition 2 and Proposition 3 of
\cite{Ander}, respectively. Taking a constant fixed source power and the
equal-phase assumption into account, the DF rate can be rewritten
as
\begin{align}\nonumber
&\frac12\max\limits_{0\leq\rho\leq1}\min\big\{
C(|h_{21}|^2P_1)+C\big(\overline{\rho^2}|h_{31}|^2P_1\big),C(|h_{31}|^2P_1)\\
\nonumber&
~~~+C(|h_{31}|^2P_1+2|h_{32}|^2P_2+2\rho\sqrt{2|h_{31}|^2P_1|h_{32}|^2P_2})\big\},
\end{align}
where $\overline{\rho^2}=1-\rho^2$; $C(x)\triangleq\log_2(1+x)$ represents
the Shannon formula for complex based model \cite{Thomas}; $\rho$
represents the correlation coefficient of $X_1^{(1)}$ and
$X_2^{(2)}$. Similarly, the CF rate can be expressed as
\begin{align}\nonumber \frac12
C(h_{31}^2P_1)+\frac12
C\Big(h_{31}^2P_1+\frac{2h_{21}^2P_1h_{32}^2P_2}{1+h_{21}^2P_1+h_{31}^2P_1+2h_{32}^2P_2}\Big).
\end{align}
Define
\begin{align}\nonumber
S_1\triangleq|h_{31}|^2P_1;~t\triangleq|h_{31}|^2/|h_{31}|^2.
\end{align}
Then we can further express the
DF rate and CF rate as functions of $S_2$:
\begin{align}\nonumber
R_{DF}(S_2)=&\frac12\max\limits_{0\leq\rho\leq1}\min\big\{
C(tS_1)+C(\overline{\rho^2}S_1),\\
\label{e:DFLB}
&~~~~~~~~C(S_1)+C(S_1+S_2+2\rho\sqrt{S_1S_2})\big\},\\
\label{e:CFLB}R_{CF}(S_2)=&\frac{C(S_1)}2+\frac12
C\Big(S_1+\frac{tS_1S_2}{1+(t+1)S_1+S_2}\Big).
\end{align}

\begin{thm}
Both the DF rate $R_{DF}(S_2)$ and CF rate $R_{CF}(S_2)$ are concave
w.r.t. $S_2$.
\end{thm}

\emph{Proof:}~First, we analyze the concavity of $R_{DF}(S_2)$. In
$R_{DF}(S_2)$, the optimal $\rho$ can be found by considering
\begin{align}\nonumber
C(tS_1)+C(\overline{\rho^2}S_1)=C(S_1)+C(S_1+S_2+2\rho\sqrt{S_1S_2})
\end{align}
which results in
\begin{align}\label{e:rho}
\rho=\frac{\sqrt{S_2+\eta(tS_1-S_1-S_2)}-\sqrt{S_2}}{\eta\sqrt{S_1}}\triangleq\rho^*
\end{align}
where $\eta\triangleq(1+tS_1)/(1+S_1)$. Note that the first and the
second terms in the minimum operation of \eqref{e:DFLB} are
monotonically decreasing and increasing w.r.t. $\rho$, $\rho\in[0,1]$. Then we have
\begin{multline}
R_{DF}(S_2)=\frac12 C(S_1)+\frac12\min\big\{C(tS_1),\\
\label{e:dflb} C\big((\sqrt{S_1}+\sqrt{S_2})^2\big),C(S_1+S_2+2\rho^*\sqrt{S_1S_2})\big\}
\end{multline}

As minimum operation is a concavity-preserving \cite{Boyd}, to show
$R_{DF}(S_2)$ is concave, we only need to show all the three terms
in the minimum operation are concave. The concavity of $C(tS_1)$ is
trivial. Note that logarithmic function is concave. According to the
composition law of concavity, to show the rest two terms in
\eqref{e:dflb} are concave, it is equivalent to show
$(\sqrt{S_1}+\sqrt{S_2})^2$ and
\begin{align}\nonumber
g_d(S_2)\triangleq S_1+S_2+2\rho^*\sqrt{S_1S_2}
\end{align}
are concave w.r.t. $S_2$ \cite{Boyd}.
On the one hand, it is not hard to see
\begin{align}\nonumber
\frac{\dif
[(\sqrt{S_1}+\sqrt{S_2})^2]}{\dif
S_2}=-\frac{\sqrt{S_1}}{2S_2\sqrt{S_2}}<0
\end{align}
 which implies the
concavity of $(\sqrt{S_1}+\sqrt{S_2})^2$. On the other hand, with
some manipulations, one has
\begin{align}\nonumber
g_d(S_2)=&S_1+S_2+\frac{2}{\eta}\Big(\sqrt{S_2^2+\eta(tS_1-S_1-S_2)S_2}-S_2\Big),\\
g_d^{\prime}(S_2)=&1-\frac2\eta+\frac{2(1-\eta)S_2+\eta(tS_1-S_1)}{\eta\sqrt{S_2^2+\eta(tS_1-S_1-S_2)S_2}},\\
\nonumber g_d^{\prime\prime}(S_2)=&\frac{2(1-\eta)}{\eta\sqrt{S_2^2+\eta(tS_1-S_1-S_2)S_2}}\\
\nonumber&-
\frac{[2(1-\eta)S_2+\eta(tS_1-S_1)]^2}{2\eta[S_2^2+\eta(tS_1-S_1-S_2)S_2]^{\frac32}}\\
\nonumber=&\frac{-[\eta(tS_1-S_1)]^2}{2\eta[S_2^2+\eta(tS_1-S_1-S_2)S_2]^{\frac32}}\leq0.
\end{align}
That is, $g_d(S_2)$ is also concave. This implies the concavity of
the DF rate $R_{DF}(S_2)$.

Next, we show that $R_{CF}(S_2)$ is concave. Let us define
\begin{align}\nonumber
g_c(S_2)\triangleq 1+S_1+\frac{tS_1S_2}{1+(t+1)S_1+S_2}.
\end{align}
Then
\begin{align}\nonumber
R_{CF}(S_2)=\frac12C(S_1)+\frac12 \log_2[g_c(S_2)].
\end{align}
According
to the composition law of concavity  \cite{Boyd}, it is equivalent
to show that $g_c(S_2)$ is concave. Note that
\begin{align}\label{e:gcderivation}
g_c^{\prime}(S_2)=&\frac{tS_1[1+(t+1)S_1]}{[1+(t+1)S_1+S_2]^2},\\
\nonumber
g_c^{\prime\prime}(S_2)=&\frac{-2tS_1[1+(t+1)S_1]}{[1+(t+1)S_1+S_2]^3}<0.
\end{align}
Then the CF rate $R_{CF}(S_2)$ is concave w.r.t. $S_2$.

\subsection{Optimal power allocation corresponding to DF and CF}

As both the DF rate and CF rate are concave w.r.t. $S_2$, one can
derive the optimal  relay power allocation according to the
necessary condition \eqref{e:optcond2} by well-defining the reverse
function of $R_{DF}^{\prime}(S_2)$ and $R_{CF}^{\prime}(S_2)$.

For $\sigma\in\{DF,CF\}$, let us denote the reverse function of
$R_{\sigma}^{\prime}(S_2)$ by $T_{\sigma}(\nu)$. We have the
following theorem on the power allocation corresponding to DF and
CF.

\begin{thm}\label{thm:DFCFPA}
For $\sigma\in\{DF,CF\}$, the optimal relay power allocation
corresponding to protocol $\sigma$ is given by
\begin{align}\label{e:S2solution}
P_{2,\sigma}^\star(\vec{h})=S_{2,\sigma}^\star(\vec{h})/(2|h_{32}|^2)\triangleq\frac1{2|h_{32}|^2}T_{\sigma}\Big(\frac{\mu^\star_{\sigma}}{2|h_{32}|^2}\Big)
\end{align}
where $\mu_{\sigma}^\star$ satisfies \eqref{e:cond2}.
\end{thm}

\emph{Proof:~} The solution can be naturally  derived from
\eqref{e:optcond2} by regarding it as an equation of $S_2$. Further
noting that
$S_2(\vec{h})=2|h_{32}|^2P_2(\vec{h})$, 
we can express the optimal power allocation corresponding to
strategy $\sigma$ as \eqref{e:S2solution}. $\hfill\blacksquare$

Next, let us analyze $T_{CF}(\nu)$ and $T_{DF}(\nu)$ in detail.

First, it is straightforward to see
\begin{align}\nonumber
R_{CF}^{\prime}(S_2)=&\frac{g_c^{\prime}(S_2)}{2g_c(S_2)\ln2}=\frac{t[1+(t+1)S_1]}{2[1+(t+1)S_1+S_2]^2g_c(S_2)\ln2}.
\end{align}
In fact, $[1+(t+1)S_1+S_2]^2g_c(S_2)$ is a quadratic polynomial
w.r.t $S_2$.  Then $T_{CF}(\nu)$ can be expressed as the positive
solution of quadratic equation in $S_2$:
\begin{align}\nonumber
[1+(t+1)S_1+S_2]^2g_c(S_2)=\frac{t[1+(t+1)S_1]}{2\nu\ln2}.
\end{align}

According to \eqref{e:dflb}, $R_{DF}(S_2)$ is a continuous piecewise
function.  Due to that $R_{DF}^{\prime}(S_2)$ is not continuous,
analysis on $T_{DF}(\nu)$ becomes complicated. By comparing the
three terms in \eqref{e:dflb}, it is not hard to rewrite $R_{DF}(S_2)$
in a piecewise form as
\begin{align}\nonumber
R_{DF}&(S_2)=\frac12 C(S_1)+\\
\label{e:Rdfpiece}&\begin{cases}
\frac12C\big((\sqrt{S_1}+\sqrt{S_2})^2\big),&\rho^*>1 \\
\frac12C(S_1+S_2+2\rho^*\sqrt{S_1S_2}),&\rho^*\in[0,1]\\
\frac12C(tS_1), &\rho^*<0
\end{cases}
\end{align}

In fact, if $t>1$, then $\rho^*>1$ is equivalent to
\begin{align}\nonumber
\sqrt{S_2+\eta(tS_1-S_1-S_2)}-\sqrt{S_2}>\eta\sqrt{S_1}
\end{align}
That is,
\begin{align}\nonumber
tS_1-S_1-S_2>\eta S_1+2\sqrt{S_1S_2}.
\end{align}
If $t-\eta>1$, or equivalently $t>S_1+2$, it arrives that
\begin{align}\label{e:f1S1}
S_2<S_1(\sqrt{t-\eta}-1)^2\triangleq f_1(S_1).
\end{align}
Similarly, if $t>1$, then $\rho^*<0$ is equivalent to
\begin{align}\label{e:f2S1}
S_2>(t-1)S_1\triangleq f_2(S_1).
\end{align}

After some manipulations, we have
\begin{align}\nonumber
&R_{DF}^{\prime}(S_2)=\\
\label{e:Rdfpiece}&\begin{cases}
\frac{\sqrt{S_1/S_2}+1}{2(1+S_1+S_2+2\sqrt{S_1S_2})\ln2},&0<S_2<f_1(S_1), \\
\frac{g_d^{\prime}(S_2)}{2(1+S_1+S_2+2\rho^*\sqrt{S_1S_2})\ln2}.&f_1(S_1)<S_2<f_2(S_1),\\
0, &S_2>f_2(S_1)
\end{cases}
\end{align}

It is easy
to verify that $g_d^{\prime}[f_2(S_1)]=0$. Hence,
$R_{DF}^{\prime}[f_2^+(S_1)]=R_{DF}^{\prime}[f_2^-(S_1)]=0$ and
$R_{DF}^{\prime}[f_2(S_1)]=0$. However, with some manipulations, one
can show that $g_d^{\prime}[f_1(S_1)]<\sqrt{S_1/f_1(S_1)}+1$.
Accordingly,
$R_{DF}^{\prime}[f_1^+(S_1)]<R_{DF}^{\prime}[f_1^-(S_1)]$ and
$R_{DF}^{\prime}[f_1(S_1)]$ does not exist. According to these
analysis, we can define the reverse function of
$R_{DF}^{\prime}(S_2)$ as follows.

\begin{itemize}
\item If
\begin{align*}
0\leq\nu<\frac{g_d^{\prime}[f_1(S_1)]}{2(1+[\sqrt{S_1}+\sqrt{f_1(S_1)}]^2)\ln2},
\end{align*}
then $T_{DF}(\nu)$ is set to the solution of equation in
$S_2$:
\begin{align*}
g_d^{\prime}(S_2)=2(1+S_1+S_2+2\rho^*\sqrt{S_1S_2})\nu\ln2.
\end{align*}
\item If
\begin{align*}
\nu>\frac{\sqrt{S_1/f_1(S_1)}+1}{2(1+[\sqrt{S_1}+\sqrt{f_1(S_1)}]^2)\ln2},
\end{align*}
then $T_{DF}(\nu)$ is set to the solution of equation regarding of
$S_2$
\begin{align*}
\sqrt{S_1/S_2}+1=2(1+S_1+S_2+2\sqrt{S_1S_2})\nu\ln2.
\end{align*}
\item Otherwise, $T_{DF}(\nu)$ is set to $f_1(S_1)$.
\end{itemize}

With the definition of $T_{\sigma}(\nu)~(\sigma\in\{DF,CF\})$, one
can search for $\nu_{\sigma}^{\star}$ in Theorem \eqref{thm:DFCFPA}.
This not only helps implementation for power allocation but also
provides clues for analyzing the power allocation in combining DF
and CF protocols.

\section{Optimal power allocation based on selecting the better one
between DF and CF}

As stated previously, the
protocol with selecting a better rate
between DF and CF can be expressed as
\begin{align}\nonumber
R(S_2)\triangleq\max\{R_{DF}(S_2),R_{CF}(S_2)\}.
\end{align}
Then, in a static relay channel, $R(S_2)$ is achievable by switching
to the better one between DF and CF protocols according to the
channel gains.

The selection is significant by noting that if $t>1$, neither DF nor
CF outperforms the other for all the relay power. Define
\begin{align}\label{e:fS1}
f(S_1)\triangleq(t-1)\big(1+(t+1)S_1\big).
\end{align}
 One can easily verify that,
if $S_2>f(S_1)$, then $R_{CF}(S_2)>R_{DF}(S_2)$ and if $S_2<f(S_1)$,
then $R_{CF}(S_2)>R_{DF}(S_2)$. Accordingly, we have
\begin{align}\label{e:CFDF}
R(S_2)=\begin{cases}R_{DF}(S_2),&S_2\leq f(S_1)\\
R_{CF}(S_2),&S_2>f(S_1).
\end{cases}
\end{align}

It is noted that
\begin{multline}\nonumber
R^{\prime}[f^+(S_1)]=R_{CF}^{\prime}[f(S_1)]\\
>0=R_{DF}^{\prime}[f(S_1)]=R^{\prime}[f^-(S_1)].
\end{multline}

Therefore, $R(S_2)$ is not concave w.r.t. $S_2$ anymore. 
In a fading HDRC, we cannot use \eqref{e:optcond2} to find the
optimal power allocation corresponding to $R(S_2)$ as what we have
done for the case using DF/CF protocol. To find some possible
solutions, let us introduce the concave envelops of $R(S_2)$,
$\overline{R}(S_2)$. In general, $\overline{R}(S_2)\geq R(S_2)$ for
all $S_2\geq0$ and $\overline{R}(S_2)$ is concave. Particularly, for
any concave function $\tilde{R}(S_2)$ satisfying $\tilde{R}(S_2)\geq
R(S_2)$ for all $S_2\geq0$, one has
$\tilde{R}(S_2)\geq\overline{R}(S_2)$.

As both $R_{DF}(S_2)$ and $R_{CF}(S_2)$ are concave and
monotonically increasing functions of $S_2$, it is easy to deduce
that $\overline{R}(S_2)$ is made up of three parts: a curve coincident with
$R_{DF}(S_2)$, a line segment connecting two points and another
curve coincident with $R_{CF}(S_2)$. In particular, the two end
points of the line segment should be located on $R_{DF}(S_2)$  and
$R_{CF}(S_2)$, respectively. What's more, if
$R_{\sigma}(S_2)~(\sigma\in\{DF,CF\})$ is smooth at the end point,
the line segment should be tangent with $R_{\sigma}(S_2)$. Assume
the two end points of the line segment are $(S_d,R_{DF}(S_d))$ and
$(S_c,R_{CF}(S_c))$, respectively where $S_d<S_c$. Then, the slope
of the line segment is given by
\begin{align}\nonumber
K\triangleq\frac{R_{CF}(S_c)-R_{DF}(S_d)}{S_c-S_d}.
\end{align}
Besides, it
also has $\overline{R}^{\prime}(S_c^+)\leq
K\leq\overline{R}^{\prime}(S_d^-)$. Accordingly, we can express
$\overline{R}(S_2)$ as
\begin{align*}
\overline{R}(S_2)=
\begin{cases}R_{DF}(S_2),&0<S_2\leq S_d\\
R_{DF}(S_d)+K(S_2-S_d),& S_d< S_2\leq S_c\\
R_{CF}(S_2),&S_2>S_c.
\end{cases}
\end{align*}
Naturally, the derivation of $\overline{R}(S_2)$ can be expressed as
\begin{align}\label{e:Rbarprime}
\overline{R}^\prime(S_2)=\begin{cases}R_{DF}^\prime(S_2),&0<S_2<S_d\\
K,& S_d< S_2< S_c\\
R_{CF}^\prime(S_2),&S_2>S_c.
\end{cases}
\end{align}
Let us denote the reverse function of $\overline{R}^\prime(S_2)$ by
$T(\nu)$. If $S_d<S_2<S_c$, then $\overline{R}(S_2)=K$ always holds. Therefore, one can define uncountable version of $T(\nu)$. Similar to the definition of $T_{DF}(\nu)$, let us define
$T(\nu)$ as follows.
\begin{itemize}
\item If there is a non-empty set $\mathcal{S}$ satisfying
that for each $S_2\in\mathcal{S}$,
$\overline{R}^{\prime}(S_2^+)\leq\nu\leq\overline{R}^{\prime}(S_2^-)$
holds, then $T(\nu)$ is set to the infimum of $\mathcal{S}$.
\item Otherwise, set $T(\nu)=0$.
\end{itemize}

It can be readily seen from the definition of $T(\nu)$ that the smallest receiver side SNR of the relay-destination link, or equivalently, the least relay power, is
selected among those satisfying the necessary condition \eqref{e:optcond2}. In fact, for $T(\nu)<S_d$ and $T(\nu)>S_c$, this definition of $T(\nu)$ is the same as that of $T_{DF}(\nu)$ and $T_{CF}(\nu)$, respectively. This specific definition of $T(\nu)$ induces a near optimal solution for power allocation based on $R(S_2)$. We summarize the result in following theorem.

\begin{thm}\label{thm:PA}
Given
\begin{align}\label{e:S2solution}
P_{2}^\star(\vec{h})=S_{2}^\star(\vec{h})/(2|h_{32}|^2)\triangleq\frac1{2|h_{32}|^2}T\Big(\frac{\mu^\star}{2|h_{32}|^2}\Big)
\end{align}
where $\mu^\star$ satisfies \eqref{e:cond2}. Then $P_2^{\star}(\vec{h})$ is a near optimal solution for relay power allocation problem
$\mathcal{P}$ based on $R(S_2)$ which is achieved by selecting the better protocol between
DF and CF.
\end{thm}

\emph{Proof: } Similar to what we have done for $R_{DF}(S_2)$ and
$R_{CF}(S_2)$, if we use $\overline{R}(S_2)$ as the static rate
performance of the system, we can get an optimal power allocation
$P_{2}^\star(\vec{h})$ following from \eqref{e:optcond2} and
$T(\nu)$. $\hfill\blacksquare$

Interestingly, the obtained average rate corresponding to
$\overline{R}(S_2(\vec{h}))$ also can be achieved by
${R}(S_2(\vec{h}))$ since
$\overline{R}(S_2(\vec{h}))=R(S_2(\vec{h}))$ holds for the solution
$S_2(\vec{h})$. This can be verified by noting that
$\overline{R}(S_2)>R(S_2)$ holds if and only if $S_d< S_2< S_c$. In
fact, for any $S_2$ satisfying $S_d\leq S_2\leq S_c$, it has
$\overline{R}^{\prime}(S_2^+)\leq K\leq\overline{R}^{\prime}(S_2^-)$
and $T(K)=S_d$.

Due to the fact that $\overline{R}(S_2)\geq R(S_2)$ holds in
general, the obtained power allocation can guarantee a near-optimal
rate performance.$\hfill\square$

\section{Conclusion}
We investigated relay power allocation over channel state in fading
HDRC based on both DF protocol and CF protocol. By proving the
concavity of the DF rate and CF rate, a parameterized form solution
for the optimal power allocation has been presented. Furthermore, we
considered a hybrid DF and CF protocol and introduced an auxiliary
function which helped find a near optimal solution of the
corresponding relay power allocation problem. 
\ifCLASSOPTIONcaptionsoff
  \newpage
\fi



%

\end{document}